# Multi-dark-state resonances in cold multi-Zeeman-sublevel atoms

Bo Wang,[1] Yanxu Han,[1] Jintao Xiao,[1] Xudong Yang,[1] Changde Xie,[1] Hai Wang,[1,*] and Min Xiao[1,2]

[1] *The State Key Laboratory of Quantum Optics and Quantum Optics Devices, Institute of Opto-Electronics, Shanxi University, Taiyuan, 030006, People's Republic of China*

[2] *Department of Physics, University of Arkansas, Fayetteville, Arkansas 72701, USA*

We present our experimental and theoretical studies of multi-dark-state resonances (MDSRs) generated in a unique cold rubidium atomic system with only one coupling laser beam. Such MDSRs are caused by different transition strengths of the strong coupling beam connecting different Zeeman sublevels. Controlling the transparency windows in such electromagnetically induced transparency system can have potential applications in multi-wavelength optical communication and quantum information processing.

Since the early demonstrations of the electromagnetically induced transparency (EIT) phenomenon in various three-level atomic systems,[1-4] many new EIT-related phenomena have been discovered and studied.[5] One of such interesting effects is dual-EIT or dual dark-state resonances studied in two coupled three-level EIT systems (such as tripod, Y-type, inverted Y-type, or other four-level atomic systems) [6-8] or by perturbing one of the lower states in the three-level EIT systems (through using one additional optical or microwave field connecting to the fourth auxiliary energy level).[9-12] In those dual EIT systems, two transparency windows are created, which can be used to allow transmissions of two probe beams simultaneous at two different wavelengths. The typical conditions for observing such dual dark-state resonances are four energy levels and three optical (or two optical and one microwave) fields.

Although two-photon Doppler-free condition allows the use of cw, low power diode lasers to eliminate first-order Doppler effect and made it possible to observe EIT with a relatively low coupling power,[3,4] the use of cold atoms can further reduce the power requirement for the coupling beam and have narrow EIT windows.[13-15] Dual dark-state phenomenon was also observed in cold atoms by perturbing the lower state for the coupling transition in a three-level Λ-type system with a microwave field.[11] Bichromatic EIT was studied in cold atoms by using two coupling laser beams and multiple EIT windows (although small) were observed.[16] Also, effects of degenerate Zeeman sublevels and various polarization configurations on EIT were studied in detail in cold atomic samples.[14]

In this letter, we propose and experimentally demonstrate a new scheme to generate multi-dark-state resonances in a three-level (with multi-Zeeman sub-levels) atomic system with only two laser fields. First, let's consider two atomic systems in $^{87}$Rb atoms, as shown in Fig.1. In Fig.1(a), the linearly polarized coupling beam $E_c$ (frequency $\omega_c$) drives transition $5S_{1/2}$, F=2 to $5P_{1/2}$, F'=1. Since the transition strengths between different Zeeman levels are very close,[17] the Rabi frequencies (defined as $\Omega_{cij} = \dfrac{\mu_{ij} \cdot E_c}{\hbar}$, where $\mu_{ij}$ is the transition dipole moment with i, j denoting Zeeman sublevels) for different coupling transitions are almost equal. The linearly-polarized (orthogonal or parallel to the polarization of the coupling beam) probe beam will have a simple single-peak EIT as studied extensively before.[15] However, when one chooses the up energy level to be F'=2 instead F'=1, as shown in Fig.1 (b), the situation changes dramatically. Due to the large differences in the transition strengths for the coupling beam connecting different Zeeman sublevels, the Rabi frequencies for different Zeeman transitions are very different. For example, the effective Rabi frequencies for transitions $5S_{1/2}$, F=2, m=±2 to $5P_{1/2}$, F'=2, m=±2 are twice the values for the transitions F=2, m=±1 to F'=2, m=±1, respectively. Transition strength from F=2, m=0 to F'=2, m=0 is zero.[17] The effective Rabi splittings (in the dressed-state picture) for the up energy levels (F'=2) have five split energy levels to be $-\dfrac{\Omega_{c22}}{2}$, $-\frac{\Omega_{c11}}{2}$, 0, $+\frac{\Omega_{c11}}{2}$, $+\frac{\Omega_{c22}}{2}$, respectively. In this particular system, $\Omega_{c22}=2\Omega_{c11}$. As the linearly-polarized (perpendicular to the polarization of the coupling beam) probe field is scanned through these split energy levels, an absorption peak appears at the center frequency for the F'=2, m=0 transition (with probe fields $\Omega_{p3}^{-}$ connecting states $|a_3\rangle$ and $|c_3\rangle$ and $\Omega_{p1}^{+}$ connecting states $|a_1\rangle$ and $|c_3\rangle$) due to lack of EIT (as shown in curve (i) of Fig. 2). The coupling field $\Omega_{c11}$ (connecting states $|b_2\rangle$ and $|c_2\rangle$) and $\Omega_{p2}^{-}$ (connecting states $|a_2\rangle$ and $|c_2\rangle$) form one EIT system, which is degenerate with another EIT system (formed by states $|a_2\rangle$, $|c_4\rangle$ and $|b_4\rangle$, with coupling field $\Omega_{c11}$ and probe field $\Omega_{p2}^{+}$). Such EIT system shows an EIT window with two absorption peaks at the positions of the dressed states, as shown in curve (ii) of Fig. 2. Similarly, coupling field $\Omega_{c22}$ (connecting states $|b_1\rangle$ and $|c_1\rangle$, as well as $|b_5\rangle$ and $|c_5\rangle$) together with probe fields $\Omega_{p1}^{-}$ (connecting states $|a_1\rangle$ and $|c_1\rangle$) and $\Omega_{p3}^{+}$ (connecting states $|a_3\rangle$ and $|c_5\rangle$) provide a larger EIT window, as shown in the curve (iii) of Fig.2. When adding these three curves in Fig.2 together, the central absorption peak and two absorption peaks in the first EIT curve cut the broad EIT window into four narrow EIT windows forming four separate dark-state resonances. Such multi-dark resonances are created by only one coupling laser beam, and the splitting frequency differences between these dark-state resonances are determined by the coupling field amplitude $E_c$. As the coupling field amplitude changes, the spaces and widths of these dark–state peaks will change accordingly.

    We performed the experiment using $^{87}$Rb atoms in a vapor-cell magneto-optical trap (MOT). Pressure of the vapor cell is ~$10^{-9}$ Torr, and two anti-helmholtz coils generate the quadrupole magnetic field with an axial gradient of 10 G/cm. An extended cavity diode laser supplies three retroreflected, circularly polarized trapping beams in three perpendicular spatial directions, and the power of each trapping beam is about 9 mW. The frequency of the trapping laser is tuned to 13 MHz below the $5S_{1/2}$,

F=2→5P$_{3/2}$, F'=3 transition. Another current controlled diode laser (with a power of 13mW) is used as the repumping laser. The frequency of this laser is lock to the 5S$_{1/2}$, F=1→5P$_{3/2}$, F'=1 transition. The diameters of the trapping beam and the repumping beam are about 1 cm and 1.5 cm, respectively. The trapped $^{87}$Rb atom cloud is ~2 mm in diameter and contains ~10$^9$ atoms at the temperature ~200μK. The probe and coupling lasers are provided by two extended cavity diode lasers with linewidth of ~1.5 MHz. The probe laser frequency is scanned at a speed of 1.3 kHz/s across the 5S$_{1/2}$, F=1→5P$_{1/2}$, F'=2 transition. The frequency of coupling laser is locked to the 5S$_{1/2}$, F=2→5P$_{1/2}$, F'=2 transition. We choose the probe beam to be linearly polarized in p direction (parallel to the optical table) and the coupling beam to be linearly polarized in s direction (perpendicular to the optical table). The propagation directions between the probe beam and the coupling beam have an angle ~20° and the diameters of the coupling beam and probe beam are ~2.0 mm and ~0.5 mm, respectively. Two beams overlap at the center of the cold atom cloud. During the experiment, the on-off sequence of the trapping, repumping, coupling and probe beams is controlled by four acousto-optical modulators (AOM). The experiment is cycled at 10 Hz. In each cycle, first 99 ms is used for the cooling and trapping of the $^{87}$Rb atoms. We switch on the coupling beam at the same time as we turn off the MOT (trapping and repumping beams, as well as the magnetic field). 0.1 ms later, the probe beam is turned on and the transmitted probe beam from the cold atoms is detected by an avalance photodiode. Through out the experiment a small magnetic field (~0.04 G) in s direction defines the quantization axis of the atoms. So the coupling beam couples the Zeeman levels of Δm=0, while the probe beam couples the Zeeman levels of Δm=±1, with the left- and right-circularly polarized components, as shown in Fig. 1(b).

Obviously, the coupling beam amplitude will greatly change the EIT windows. Figure 3(a) plots the experimental results of probe transmission as a function of probe frequency detuning, which probes the dark-state structure of this system. When the coupling beam is relatively weak ($\Omega_{c22}$=14 MHz, with the probe beam Rabi frequency $\Omega_p$=2 MHz), the Zeeman sublevels are near degenerate, which gives a single EIT peak as shown in Fig. 3(a1). The absorption at the transition from 5S$_{1/2}$, F=1, m=0→5P$_{1/2}$,

F'=2, m=0 reduces the EIT effect and makes the total EIT dip smaller. As the coupling field amplitude increases to a larger value, several EIT windows start to appear at different probe frequencies with an absorption peak at the middle, as shown in Fig. 3(a2). Further increase of coupling power makes the split EIT windows more clear as given in Fig. 3(a3) and (a4).

The total probe field susceptibility is $\chi = \chi_1 + \chi_2 + \chi_3$, where $\chi_i$ (i=1,2,3) are the sums of the left- and right-circularly-polarized probe field susceptibilities coming from contributions of the transitions $|a_i\rangle \to |c_i\rangle$ and $|a_i\rangle \to |c_{i+2}\rangle$, respectively. These susceptibilities can be expressed as

$$\chi_1 = -\frac{N}{\hbar\varepsilon_0}\left[\frac{|\mu_{c1,a1}|^2}{\Omega_{p1}^-}\rho_{a1,c1} + \frac{|\mu_{c5,a3}|^2}{\Omega_{p3}^+}\rho_{a3,c5}\right],$$

$$\chi_2 = -\frac{N}{\hbar\varepsilon_0}\left[\frac{|\mu_{c2,a2}|^2}{\Omega_{p2}^-}\rho_{a2,c2} + \frac{|\mu_{c4,a2}|^2}{\Omega_{p2}^+}\rho_{a2,c4}\right],$$

$$\chi_3 = -\frac{N}{\hbar\varepsilon_0}\left[\frac{|\mu_{c3,a3}|^2}{\Omega_{p3}^-}\rho_{a3,c3} + \frac{|\mu_{c3,a1}|^2}{\Omega_{p1}^+}\rho_{a1,c3}\right].$$

$\rho_{ai,ci}$ and $\rho_{ai,ci+2}$ can be calculated by solving the density-matrix equations involving all Zeeman sublevels in the system.[18,19] N is the atomic density.

When the coupling beam is weak ($\Omega_{c22}$=14 MHz, which is still much stronger than the probe beam), the Rabi splittings for different Zeeman sublevels are very small, so the total probe susceptibility $\chi$ shows a single-peak EIT profile (Fig. 3(b1)). As the coupling field amplitude increases to a larger value ($\Omega_{c22}$=31 MHz), one can start to see the split of the EIT window into four small EIT peaks with absorption at the center frequency (Fig. 3(b2)). When the coupling beam is further increased (corresponding to $\Omega_{c22}$=56 MHz and 78 MHz, respectively), the splitting gets larger and the EIT dips become more pronounced as shown in Figs. 3(b3), and (b4). Curves in Fig.3(b) use similar parameters as the ones used in the experimental measurements and the agreements between these experimentally measured curves and theoretically calculated ones are excellent.

The frequencies of these multi-EIT windows (or dark states) can be controlled by the coupling beam power, which allows transmissions of the probe beam at different frequencies. Such tunable EIT windows can be used for multi-channel optical communication or quantum information processing. With such sharp EIT windows, probe photons at different frequencies can be slowed down simultaneously. Figure 4 depicts dispersion properties for the cases of small (Fig. 4(a)) and large (Fig. 4(b)) coupling beam strengths, respectively. As one can see, sharp normal dispersion appears at different frequencies and can be used to modify group velocities of the probe beams at different frequencies for multi-channel optical communication.

In conclusion, we have demonstrated multi-dark resonances in a unique multi-Zeeman-sublevel atomic system in cold atoms with only one coupling beam. The differences in coupling Rabi frequencies due to different transition strengths among different Zeeman sublevels, together with the absorption peak at m=0 transition, split the broad EIT window into four narrow EIT windows which are controllable by coupling beam power. Such interesting effect only occurs in very unique atomic systems with certain Zeeman sublevels and can only be observed in cold atoms. Such controllable multiple EIT windows can be used for multi-channel optical communication and multi-channel quantum information processing.

We acknowledge funding supports by the CNSF (# 60325414, 60578059, 60238010 and RGC60518001) in China and PCSIRT. *Corresponding author H. Wang's e-mail address is wanghai@sxu.edu.cn.

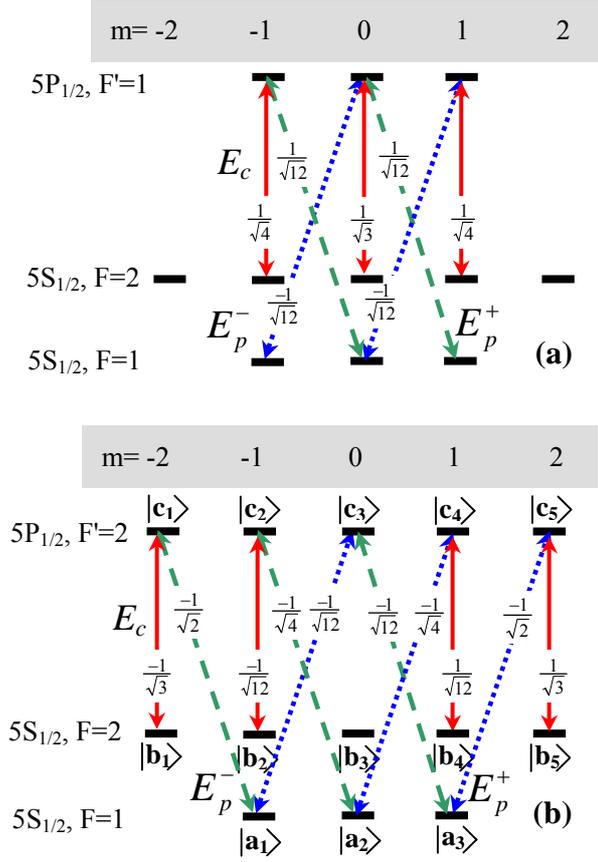

Fig.1. (Color online) Λ-type EIT subsystems formed by linearly-polarized probe and coupling fields (the polarizations of these two beams are perpendicular to each other) in D1 line of $^{87}$Rb atoms. (a) The upper energy level is $5P_{1/2}$, F'=1; (b) The upper energy level is $5P_{1/2}$, F'=2.

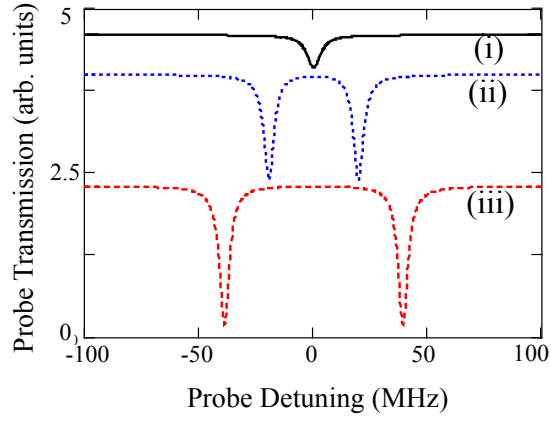

Fig.2 (Color online) The absorption spectra of the probe field coupling to the upper Zeeman Sublevel (for the system shown in Fig. 1(b)) of (i) F'=2, m=0, (ii) F'=2, m=±1, and (iii) F'=2, m=±2, respectively. Calculation parameters are $\Omega_{c22}$=78 MHz, $\Omega_p$=2 MHz, $\gamma_{ab}$=40 kHz, $\gamma_{ac}$=2.8MHz (the curves have been vertically shifted for clarity).

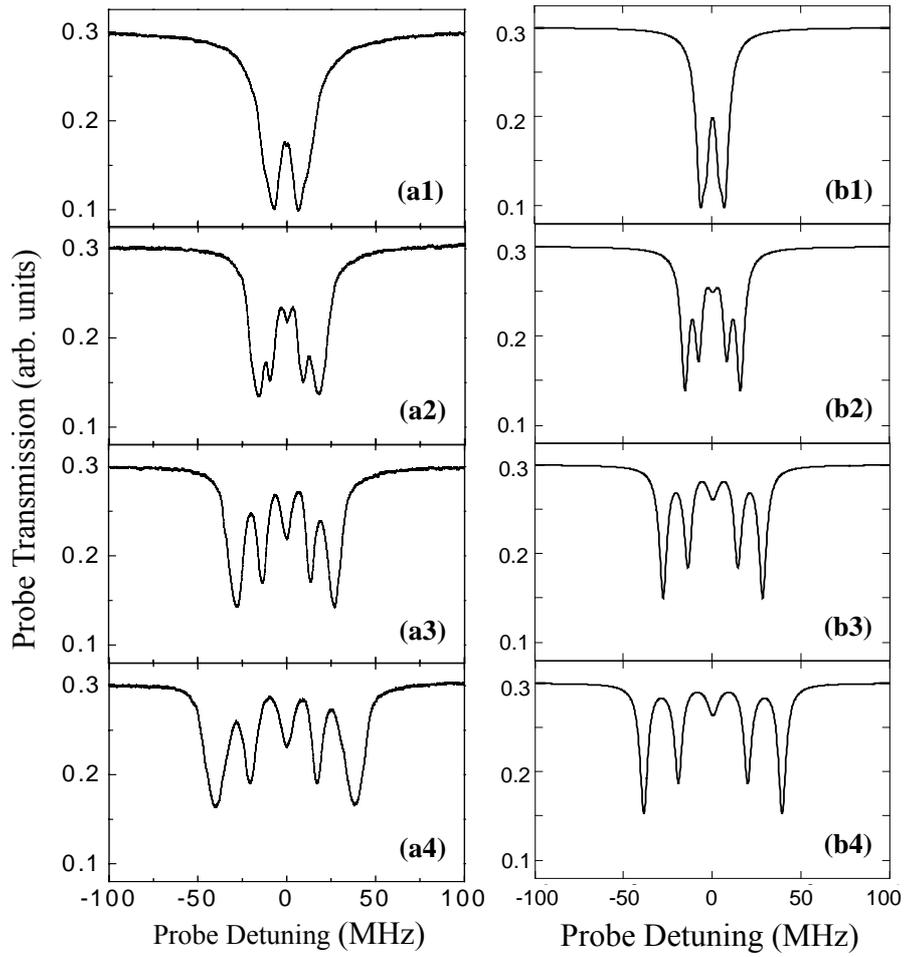

Fig.3 (a) The MDSR spectra observed in the experiment. The probe beam Rabi frequency is $\Omega_p$=2 MHz, the coupling field Rabi frequencies in (a1)-(a4) are $\Omega_{c22}$=14 MHz, 31 MHz, 56 MHz, and 78 MHz, respectively. (b) Theoretical MDSR spectra calculated with the same parameters as in (a), other parameters used in the calculation are $\gamma_{ab}$=40 kHz, $\gamma_{ac}$=2.8MHz, laser linewidths=1.5MHz, N=1×10$^{11}$ cm$^{-3}$.

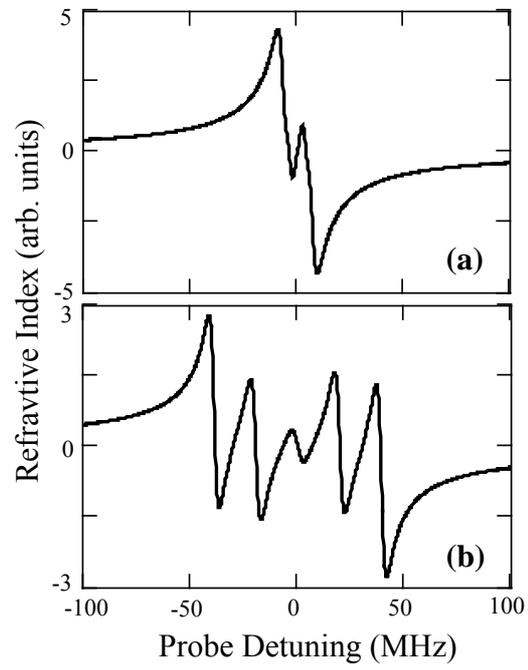

Fig.4 The calculated probe dispersion with same parameters as used in (a) Fig. 3(b1) and (b) Fig. 3(b4).